\documentclass[11pt, a4paper]{article}
\usepackage[]{graphicx}
\usepackage[]{color}
\makeatletter
\def\maxwidth{ %
  \ifdim\Gin@nat@width>\linewidth
    \linewidth
  \else
    \Gin@nat@width
  \fi
}
\makeatother

\definecolor{fgcolor}{rgb}{0.345, 0.345, 0.345}

\usepackage{framed}
\makeatletter
 {\par\unskip\endMakeFramed%
 \at@end@of@kframe}
\makeatother

\definecolor{shadecolor}{rgb}{.97, .97, .97}
\definecolor{messagecolor}{rgb}{0, 0, 0}
\definecolor{warningcolor}{rgb}{1, 0, 1}
\definecolor{errorcolor}{rgb}{1, 0, 0}

\usepackage{alltt}

\usepackage{verbatim}
\usepackage[vmargin=3cm, hmargin=2cm]{geometry}
\usepackage{url}
\usepackage{hyperref}
\usepackage{fancyhdr}
\usepackage{color}
\usepackage{amsmath, amssymb}
\usepackage{longtable}
\usepackage{lscape}
\usepackage[super, comma,numbers]{natbib}
\usepackage{xspace}
\usepackage{booktabs}
\usepackage{newfloat}
\usepackage{rotating}
\usepackage{scrextend}
\usepackage{csquotes}
\DeclareFloatingEnvironment[
    placement=!htbp
]{listing}
\usepackage{fancyvrb}
\usepackage{lmodern}
\usepackage{nameref}
\usepackage[T1]{fontenc}
\usepackage{listings}
\usepackage[font=sf, labelfont={sf}, margin=1cm]{caption}

\usepackage[latin1]{inputenc}

\IfFileExists{upquote.sty}{\usepackage{upquote}}{}
\begin{document}

\begin{center}
    \Large
    \textbf{A Bayesian time-to-event pharmacokinetic model for phase I dose-escalation trials with multiple schedules}

    \vspace{0.4cm}
    \large
    \textbf{Burak K\"ursad G\"unhan},\footnote{\textit{Department of Medical Statistics, University Medical Center G\"ottingen, G\"ottingen, Germany} \label{goettingen}} \footnote{\textit{Correspondence to: Burak K\"ursad G\"unhan; email: \texttt{burak.gunhan@med.uni-goettingen.de}}} \textbf{Sebastian Weber},\footnote{\textit{Novartis Pharma AG, Basel, Switzerland} \label{novartis}}   \textbf{Tim Friede}\footref{goettingen} 
    \vspace{0.9cm}
\end{center}

Phase I dose-escalation trials must be guided by a safety model in order to avoid exposing patients to unacceptably high risk of toxicities. Traditionally, these trials are based on one type of schedule.
In more recent practice, however, there is often a need to consider more than one
schedule, which means that in addition to the dose itself, the schedule needs to be varied in the trial. Hence, the aim is finding an 
acceptable dose-schedule combination. However, most established methods for dose-escalation trials are designed to escalate the dose only and ad-hoc choices must be made to adapt these to the more complicated setting of finding an acceptable dose-schedule combination. In this paper, we introduce a Bayesian time-to-event model which takes explicitly the dose amount and schedule into account through the use of pharmacokinetic principles. The model uses a time-varying exposure measure to account for the risk of a dose-limiting toxicity over time.
The dose-schedule decisions are informed by an escalation with overdose
control criterion. The model is formulated using interpretable 
parameters which facilitates the specification of priors. 
In a simulation study, we compared the proposed
method with an existing method. The simulation study demonstrates that 
the proposed method yields similar or better results compared to an existing method 
in terms of recommending acceptable dose-schedule combinations, yet reduces 
the number of patients enrolled in most of scenarios. The \texttt{R} and \texttt{Stan} code to implement the
proposed method is publicly available from Github (\url{https://github.com/gunhanb/TITEPK_code}).\\

\textbf{Keywords:}Phase I dose-escalation trials, multiple schedules, pharmacokinetic models, Stan

\section{Introduction} \label{sec:1}
In a phase I trial, a treatment plan 
includes the amount of drug to be given a patient, known as the 
dose, and the times
when it is given, known as the schedule. Phase I dose-escalation 
trials traditionally include 
only one schedule while varying the dose among patients. 
However, in medical practice,
there is often a need to consider different schedules, e.g. a dose given once a day or once 
a week, within a phase I trial. Many established methods\cite{10.2307/2531628,neuenschwander2015bayesian} are only designed for varying the drug amount since time is not taken into account in the models. These approaches require ad-hoc adjustments like scaling the dose to accommodate these more complex designs. However, varying treatment schedules necessitates to take time into account in the model as the pharmacokinetic properties of a drug become relevant whenever the treatment schedule is varied. 

The aim of a phase I dose-escalation trial with multiple schedules is finding an acceptable dose and schedule combination. The dose-escalation must be guided by a safety model in order to avoid exposing patients to unacceptably high risk of toxicities. What defines an acceptable dose-schedule combination depends on the drug development strategy. In oncology, the efficacy is sought to be maximized at the cost of tolerating safety events, referred to as dose-limiting toxicities (DLTs). Therefore one seeks in oncology a so-called maximum tolerated dose-schedule combination (MTC) at which an acceptable rate of DLT events is expected to occur.

Different definitions of treatment schedule have been used in the literature.\cite{Guo2016,wages2017} When the aim is to optimize the number of cycles to treat patients, the schedule is defined as the number of treatment cycles. An alternative definition is the frequency (timing) of administration within a cycle with a given total dose per cycle. Guo et al\cite{Guo2016} argued that this definition of schedule seems more relevant in practical terms, since physicians will usually continue to treat patients as long as patients appear to benefit from the treatment. A more interesting point for the physicians is how frequently the treatment should be administered. We agree with this reasoning and hence adopted this second definition of schedule in the paper.

There are different methods suggested for determining the MTC. Using the definition of a schedule as the number of treatment cycles,
Braun et al.\cite{Braun2007} developed a time-to-event model which simultaneously optimizes the dose and the schedule. Zhang and Braun\cite{Zhang2013} extended this method to incorporate adaptive variations to dose-schedule assignments within patients as the trial proceeds. Wages et al.\cite{Wages2014} introduce a dose-schedule finding design, the partial order continual reassessment method (POCRM),
which relaxes the assumption of completely ordered schedules, that is for a given dose,
DLT probabilities are completely ordered in terms of different schedules. Wages et al.\cite{Wages2014} used the second definition of the schedule.\cite{wages2017}
Furthermore, Li et al.,\cite{Li2008} Thall et al.,\cite{Thall2013} Guo et al.,\cite{Guo2016} and Cunanan and Koopmeiners\cite{Cunanan2017} suggested
dose-schedule finding methods that jointly models efficacy and toxicity in the context of dose-schedule combination designs.

We introduce an alternative model, 
a \emph{time-to-event pharmacokinetic model} 
henceforth referred as TITE-PK, that uses pharmacokinetics (PK) principles to introduce an exposure measure. Consequently, TITE-PK is
an exposure-response model\cite{pinheiro2009exposure} which usually uses
more information than a standard dose-response model, such as kinetic drug properties. Formally, a pseudo-PK model is used to define a time-varying exposure measure, which constitutes a time-varying Poisson process describing the DLT event process.
TITE-PK utilizes data on the exact treatment schedule and time-to-first DLT in a fully Bayesian model-based approach following the spirit of Cox
et al.\cite{Cox1999}
To inform dose-schedule decisions, TITE-PK uses an adapted escalation with overdose control (EWOC)\cite{SIM:SIM793} criterion. This requires that for a given dose-schedule combination, the probability for a DLT occurred within the first cycle must not exceed the maximal admissible DLT probability by a pre-specified feasibility bound. In the proposed model, PK analysis and safety analysis are not combined as is done for example by Ursino et al.\cite{BIMJ:BIMJ1759} Instead we use a pseudo-PK model in TITE-PK which can be seen as a kinetic-pharmacodynamic model (K-PD), see e.g. Jacqmin et al.\cite{Jacqmin2007} and Jacons et al.\cite{Jacobs2010}

We provide simulations comparing the performance of our proposed model to the POCRM method, which is in the spirit of the continual reassessment design (CRM).\cite{10.2307/2531628} POCRM was originally developed for drug combination trials,\cite{Wages2011} and later extended to phase I trials with multiple schedules.\cite{Wages2014} The simulation study is motivated by the Vidaza trial.\cite{Vidaza} Vidaza is a cytotoxic drug that is used for the treatment of a blood cell disease, known as myelodysplastic syndrome, that often develops into acute myelogeneos leukemia. The Vidaza trial (ClinicalTrials.gov identifier: NCT01080664) investigated four different schedules and three doses, and thus, is an example of a dose-schedule finding problem.\cite{Braun2007} The \texttt{R} and \texttt{Stan} code for the implementation of the proposed TITE-PK model is available from Github (\url{https://github.com/gunhanb/TITEPK_code}).

This manuscript is structured as follows. In Section 2, we introduce the proposed TITE-PK model. In Section 3, the performance of TITE-PK and POCRM are compared in a simulation study. We close with a discussion and a conclusion.

\section{The proposed model: TITE-PK} \label{sec:titepk}
The 
time-to-first DLT is modeled using a time-varying (non-homogeneous) 
Poisson process. A time-varying Poisson process can be defined using 
the instantaneous hazard function ($h(t)$) for a DLT occurring at time $t$. The hazard function corresponds to the probability that a patient
experiences a DLT in the time interval $(t, t + \delta t]$ given that
they did not experience a DLT until time $t$. The hazard is modeled as
a time-dependent function directly proportional to an 
exposure measure of the drug
($E(t)$) as\cite{Cox1999}
\begin{align}
  h(t) = \beta \, E(t) \label{eq:hazard_exposure}
\end{align}
where $\beta$ is the proportionality parameter to estimate. 
Here, the
exposure measure refers to the drug concentration as in an
exposure-response model,\cite{pinheiro2009exposure} and the
calculation of $E(t)$ will be explained in
Section~\ref{sec:pseudoPK}. Furthermore, if we integrate both sides of
Equation~\eqref{eq:hazard_exposure} from time 0 up to time $t$, we obtain
\begin{align}
  H(t) = \beta \, \text{AUC}_{E}(t)  \label{eq:AUC}
\end{align}
where $\text{AUC}_{E}(t)$ is the area under the curve of the exposure
measure over time and $H(t)$ is the cumulative hazard function,
respectively. From event history analysis,\citep{Kalbfleisch2002} we know that
the probability density for an event to occur at time-point $t$ is
\begin{align}
  f(t) = h(t) \, \exp(-H(t))  \label{eq:pdf}
\end{align}
and the survivor function for the event to occur past some time-point $t$ is given by
\begin{align}
  S(t) = P(T>t) = \exp(-H(t))  \label{eq:ccdf}
\end{align}
where $T$ denotes the event time.
In the following, we use $C_j$ to denote the censoring time of patient
$j$. Accordingly, TITE-PK is able to account for the partial information from subjects still in the follow-up (censored patients) like TITE-CRM.\cite{titecrm} In contrast to TITE-CRM, we restrict the follow-up period for all patients to cycle 1 only, which is a conventional approach. Thus, all patients without a DLT up to the end of
cycle 1 will be censored at the end of cycle 1, $C_{j} = t^{*}$. 
Furthermore, we denote with $\delta_{j}$ an event indicator
which is set to $0$ for censored events and $1$ for DLT events. 
The overall likelihood can be written as
\begin{align}
  L(T, C|\beta) = \prod_{j=1}^{J} f(T_{j}|\beta)^{\delta_{j}} \, S(C_{j}|\beta)^{(1 - \delta_{j})}, \label{eq:likelihood}
\end{align}
where $J$ is the total number of the patients. Now, we discuss the exposure measure of the drug. 

\subsection{Pseudo-PK model} \label{sec:pseudoPK}
The proposed exposure model in the TITE-PK model does not rely on measured drug concentration data, as this data is not routinely available in a form that it may be used directly in the model to support dose-schedule decisions in a timely manner. For this reason, PK is considered as latent variable which
we refer to \emph{pseudo}-PK. The pseudo-PK is used to account for the dosing history and the expected accumulation in exposure over time that ultimately drive pharmacological responses, including safety. The main purpose of the proposed pseudo-PK model is to account for the natural ``waxing and waning'' of exposure observed after dosing of drug. This pseudo-PK model has a \emph{central} compartment into which the drug is administered and accounts for drug elimination as a linear first order process; i.e. the elimination rate is proportional to the amount of drug in the compartment
\cite{kallen2007computational}
\begin{align}
  \frac{dC(t)}{dt} = - k_{e} \, C(t), \label{eq:1-cmt}
\end{align}
where $C(t)$ is the concentration of drug in the central compartment
and $k_{e}$ is the elimination rate constant. As the volume of the central
compartment cannot be identified for a latent \emph{pseudo}-PK, we
set it by convention to unity. 

To account for delays between the instantaneous drug concentration in the central compartment and the concentration during the pharmacodynamic effect, we use a so-called effect compartment\cite{kallen2007computational}
\begin{align}
  \frac{d C_{\text{eff}}(t)}{dt} = k_{\text{eff}} \, (C(t) - C_{\text{eff}}(t)). \label{eq:eff-cmt}
\end{align}

Here $C_{\text{eff}}(t)$ is the drug concentration in the effect
compartment and $k_{\text{eff}}$ is the PK parameter which governs
the delay between the concentration in the central compartment
($C(t)$) and the concentration in the effect compartment
($C_{\text{eff}}(t)$). Note that the solutions of 
Equation~\eqref{eq:1-cmt} and Equation~\eqref{eq:eff-cmt} are the same
up to reparametrization for a one compartment model with first order absorption, which would be one way to model oral absorption. The parameters $k_{e}$ and 
$k_{\text{eff}}$ are assumed to be known from previous PK analyses, 
for example from pre-clinical experiments. The model is conditioned on the PK parameters from previous analyses, thus uncertainty of the PK parameters in the estimation is ignored. A procedure to calculate an estimate for $k_{\text{eff}}$ using the cycle duration and the absorption rate is described in Section~\ref{settings} for the Vidaza trial.

The ordinary differential equations (ODE) \eqref{eq:1-cmt} and
\eqref{eq:eff-cmt} account for dosing over time through administration into the
central compartment. The analytical solution to the ODE system for
multiple doses is obtained through the use of the superposition
principle which holds for linear ODE systems (see e.g. \cite{bertrand2008mathematical}).
This model in principle can account for the history of any treatment schedule over time. 
In order to simplify the notation, we restrict
ourselves here to regular treatment schedules which have a
dosing frequency $f$ (in units of $1/h$), start at time $t=0h$ and use
the same dose amount $d$ for all dosing events. With these simplifications (in notation) the solution to
the above ODE system is
\begin{align}
  C_{\text{eff}}(t|d,f) = d \, \sum_{i = 0}^\infty \Theta \left(t - \frac{i}{f}\right) \frac{k_{\text{eff}}}{k_{\text{eff}} -  k_{e}} \left(e^{- k_{e}\left(t -  \frac{i}{f} \right)} - 
e^{-k_{\text{eff}}\left(t - \frac{i}{f} \right)}\right),  \label{eq:analytical}
\end{align}
where $\Theta$ denotes the Heaviside step function (or unit step
function). 

To facilitate meaningful interpretation of the parameter $\beta$ and
hence to help prior specification, the exposure measure $E(t)$ is
obtained by scaling $C_{\text{eff}}(t)$ using a reference dose-schedule combination
including a reference dose ($d^{*}$) and a reference dosing frequency
($f^{*}$) at the end of cycle 1 ($t^{*}$) such that
\begin{align*}
E(t| d, f) &= \frac{C_{\text{eff}}(t|d,f)}{\int_0^{t^{*}} C_{\text{eff}}(t|d^{*},f^{*}) \, dt} \\
\text{AUC}_{E}(t^{*}|d^{*}, f^{*}) &= \int_0^{t^{*}} E(t| d^{*}, f^{*}) \, dt = 1. 
\end{align*}

This is analogous to using a reference dose in the
Bayesian Logistic Regression Model\cite{neuenschwander2015bayesian} which is a 
two parameter version of the CRM design and uses the EWOC criterion for dose-escalation decisions.

\subsection{Informing dose-schedule decisions} \label{sec:EWOC}
To inform dose-schedule decisions, TITE-PK uses
an adapted EWOC criterion. The probability that a patient experiences
at least one DLT within the first cycle (shortly the end-of-cycle 1 
DLT probability) given the dose-schedule combination with
dose $d$ and frequency $f$, $P(T \leq t^{*}|d,f)$, is our measure of interest.

The end-of-cycle 1 DLT probabilities are classified into three categories as follows
\begin{enumerate}
  \item[(i)]\hspace{1cm}    $P(T \leq t^{*}|d,f)< 0.20$ \hspace{0.55cm} Underdosing (UD)
  \item[(ii)]   $0.20 \leq P(T \leq t^{*}|d,f) \leq 0.40$ \hspace{0.5cm} Targeted toxicity (TT)
  \item[(iii)]\hspace{1cm}  $P(T \leq t^{*}|d,f) > 0.40$ \hspace{0.55cm} Overdosing (OD)
\end{enumerate}

Dose-schedule decisions are informed using the
overdosing probability of the dose-schedule combination $d$ and $f$. 
The EWOC criterion is fulfilled, if $P(P(T \leq t^{*}|d,f) > 0.40)$ is smaller than the
pre-specified feasibility bound $a$. Among the dose-schedule combinations which fulfill the EWOC criterion, the combination which has the lowest $\text{AUC}_{E}(t^{*})$ is recommended by TITE-PK. This is analogous to recommending the lowest dose amount in the ``standard'' phase I dose-escalation methods. When $\text{AUC}_{E}(t^{*})$ of eligible combinations are exactly the same, one of the dose-schedule combinations can be chosen randomly. In this paper, we use $a = 0.25$, which was 
suggested by Babb et al.,\cite{SIM:SIM793} and also $a = 0.50$ to investigate the sensitivity of the results to the choice of $a$ 
in our simulations. The higher (lower) value of the feasibility bound make it easier 
(harder) to escalate to the next dose and schedule combinations, resulting in more (less) aggressive dose-schedule decisions.

By using the relationship between $P(T \leq t^{*}|d,f) = 1 - P(T > t^{*}|d,f)$
and combining Equation~\eqref{eq:ccdf} with~\eqref{eq:AUC} it follows that
\begin{align}
  P(T > t^{*}|d,f) = \exp(-H(t^{*}|d,f)) = & 1 - P(T \leq t^{*}|d,f) \nonumber  \\
  \Leftrightarrow \log(H(t^{*}|d,f))  = & \log(-\log(1 - P(T \leq t^{*}|d,f))) = \text{cloglog}(P(T \leq t^{*}|d,f)) \label{eq:hazard_cloglog}
\end{align}
where $\text{cloglog}(x) = \text{log}(-\text{log}(1-x))$.

Since the cumulative hazard $H(t|d,f)$ is set proportional, see
Equation~\eqref{eq:AUC}, to the area under curve of the exposure
metric $\text{AUC}_{E}(t|d,f)$ this leads to
\begin{align}
\text{cloglog}(P(T \leq t^{*} | d, f)) = \log(\beta) + \log(\text{AUC}_{E}(t^* | d, f)).
\end{align}

For the reference dose-schedule combination with dose $d^*$ and 
dosing frequency $f^*$ the AUC of the exposure measure up to the reference 
time-point is unity, $\text{AUC}_{E}(t^{*}|d^{*},f^{*})=1$, such that
$\text{cloglog}(P(T \leq t^{*} | d^*, f^*)) = \log(\beta)$ holds. This
highlights the importance of the reference dose-schedule combination
to specify the prior for the parameter $\beta$.

As is apparent from Equation~\eqref{eq:AUC}, TITE-PK assumes that the end-of-cycle 1 DLT probability is a monotonic function of the exposure measure. 
Moreover, the pseudo-PK model assumes a linear first order process (linear PK), meaning that increases in drug exposure are linearly related to increases in administered doses. Consequently, $\text{AUC}_{E}(t^{*}|d,f)$ is directly proportional to $d$. Thus, the assumption of the monotonicity of the exposure and the end-of-cycle 1 
DLT probability implies
the assumption of the monotonicity of the dose and the end-of-cycle 1 
DLT probability. However,
we will see in the simulations that the performance of the model is robust to some extent to violations of the monotonicity assumption.

\subsection{Software implementation}\label{comp}
The proposed model TITE-PK is implemented in \texttt{Stan}
\citep{JSSv076i01} via \textbf{rstan} \texttt{R} package. The corresponding
code for the implementation of the TITE-PK model is available from
Github (\url{https://github.com/gunhanb/TITEPK_code}). 
Four parallel chains of 1,000 MCMC iterations after warm-up
of 1,000 iterations are generated. Convergence diagnostics are checked
using the Gelman-Rubin statistics\cite{Gelman1992} and traceplots.

\section{Simulation study} \label{sec:oc}
In order to assess the performance of the TITE-PK
and to compare with the POCRM under different true 
dose-DLT profiles with multiple schedules, various scenarios are 
investigated in a simulation study.

\subsection{Simulation settings} \label{settings}

\begin{figure}[!htb]
  \centering
  \includegraphics[scale=0.2]{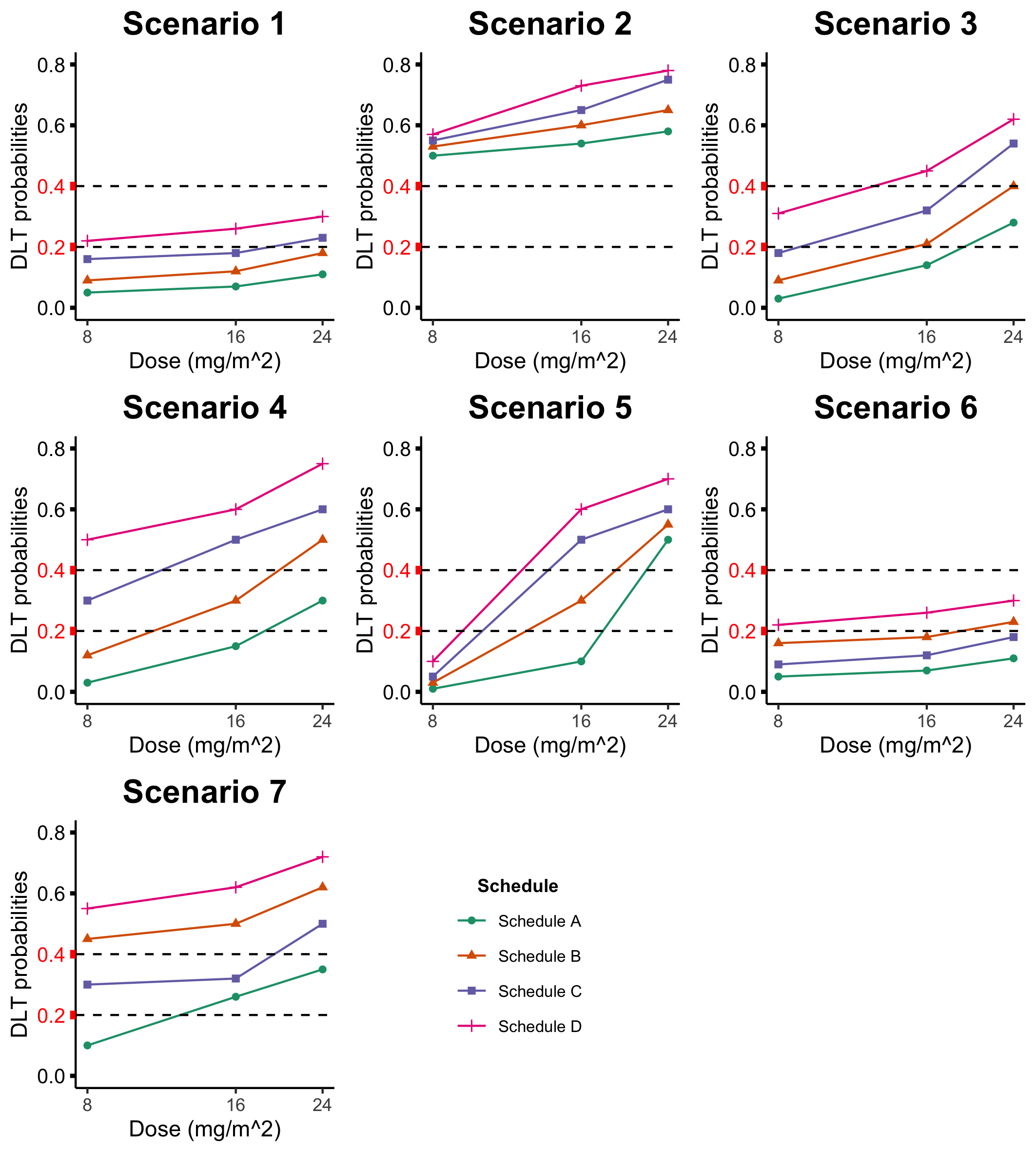}
  \caption{Toxicity scenarios for the dose-schedule combination in the simulation study. The horizontal dashed lines represent the boundaries of the targeted toxicity interval. Schedules A, B, C, and D have dosing frequency 
of 192, 96, 48, and 24 hours, respectively.} 
  \label{fig:DoseScenarios}
\end{figure}

The scenarios considered in the paper were also investigated 
by Wages et al.\cite{Wages2014} These are motivated by the Vidaza example.\cite{Vidaza} As in the 
Vidaza trial, the scenarios investigated four different schedules 
(A, B, C, D) and three doses (8, 16, 24 mg/m$^2$). In the Vidaza trial, 
different treatment schedules correspond 
to the number of cycles the drug is administered with a prespecified frequency of administration. More specifically, four schedules are 1, 2, 3, or 4 cycles, 
each with 5 days of drug administration and 25 days of 
rest. As explained in the introduction, here we use another definition of the schedule, the timing of the 
administration within the first cycle only. To mimic the 
nested schedules of the Vidaza trial, we chose the 
four schedules A, B, C, and D as dosing frequency 
of 192, 96, 48, and 24 hours, respectively. 
The cycle length is taken as 28 days ($t^*= 28$ days). The reference dose 
and the reference dosing frequency are determined using 24 
mg ($d^*= 24$ mg) and Schedule B ($f^*= 1/96$ 1/h).

\begin{table}
\centering
\caption{Toxicity scenarios for the dose-schedule combination in the simulation study. Combinations in the targeted toxicity interval (0.20 - 0.40) are in boldface. Schedules A, B, C, and D have dosing frequency 
of 192, 96, 48, and 24 hours, respectively.} 
\label{t1:scenarios}
\begin{tabular}{lllllll}
  \toprule
             & \multicolumn{6}{c}{\textbf{Doses in mg/$\text{m}^2$}} \\
             \cmidrule{2-7}
   Schedule  & 8 & 16 & 24 & 8 & 16 & 24 \\ 
  \midrule
             & \multicolumn{3}{c}{Scenario 1} & \multicolumn{3}{c}{Scenario 2} \\
                                          A & 0.05 & 0.07 & 0.11                & 0.50 & 0.54 & 0.58  \\
                                          B & 0.09 & 0.12 & 0.18                & 0.53 & 0.60 & 0.65 \\
                                          C & 0.16 & 0.18 & \textbf{0.23}   & 0.55 & 0.65 & 0.75 \\
                                          D & \textbf{0.22} & \textbf{0.26} & \textbf{0.30} & 0.57 & 0.73 & 0.78 \\ 
   
           & \multicolumn{3}{c}{Scenario 3} & \multicolumn{3}{c}{Scenario 4}\\
                                           A  & 0.03 & 0.14 & \textbf{0.28} & 0.03 & 0.15 & \textbf{0.30} \\
                                           B  & 0.09 & \textbf{0.21} & \textbf{0.40}  & 0.12 & \textbf{0.30} & 0.50\\
                                           C  & 0.18 & \textbf{0.32} & 0.54 & \textbf{0.30} & 0.50 & 0.60\\
                                           D  & \textbf{0.31} & 0.45 & 0.62 & 0.50 & 0.60 & 0.75\\

                & \multicolumn{3}{c}{Scenario 5} & \multicolumn{3}{c}{Scenario 6} \\
   A &  0.01 & 0.10 & 0.50 & 0.05 & 0.07 & 0.11 \\
   B &  0.03 & \textbf{0.30} & 0.55 & 0.16 & 0.18 & \textbf{0.23} \\
   C &  0.05 & 0.50 & 0.60  & 0.09 & 0.12 & 0.18 \\
   D &  0.10 & 0.60 & 0.70 & \textbf{0.22} & \textbf{0.26} & \textbf{0.30}\\ 
   
                &  \multicolumn{3}{c}{Scenario 7} \\
   A & 0.10 & \textbf{0.26} & \textbf{0.35}\\
   B & 0.45 & 0.50 & 0.62 \\
   C & \textbf{0.30} & \textbf{0.32} & 0.50 \\
   D & 0.55 & 0.62 & 0.72 \\ 

   \bottomrule
\end{tabular}
\end{table}

The scenarios were carefully chosen to reflect a range of clinically relevant scenarios. These are summarized in Table~\ref{t1:scenarios} and illustrated by Figure 1. Scenario 1 does not include any dose-schedule combination in the overdosing interval, whereas all combinations are in the overdosing interval in Scenario 2. Scenarios 3 and 4 are scenarios in which DLT probabilities are spread across underdosing, targeted toxicity, and overdosing intervals. Five and three dose-schedule combinations are in the targeted toxicity interval in Scenarios 3 and 4, respectively. In Scenario 5, there is only one dose-schedule combination in the targeted toxicity interval. Also, there is a heavy violation of the monotonicity assumption of DLT probabilities with increasing exposure in Scenario 5. Moreover, Scenarios 1-5 assume completely ordered schedules, that is DLT probabilities increase monotonically with schedules involving more frequent administration given the same dose. In contrast, Scenarios 6 and 7 relax this assumption, and assume partially ordered schedules. Scenario 6 correspond to Scenario 1 with DLT probabilities for Schedules B and C switched. Similarly, Scenario 7 corresponds to a scenario which spread across different intervals, but DLT probabilities for Schedules B and C are switched. Additionally, we considered more scenarios to assess the performance of TITE-PK, which are listed in Table~\ref{tab:app1}.

The elimination half-life and the absorption rate of Vidaza were reported as 4 (hours) and 2 (1/hours), respectively.\cite{PKvidaza} Thus, we specify $k_{e} = \frac{\text{log(2)}}{4}$ (1/hours). We chose this value corresponding to the mean of an adequate distribution for the parameter $k_{\text{eff}}$. Here, we chose a log-normal distribution defined by the lower and upper quantile reflecting the fastest and slowest time-scale of the experiment which is the absorption half-life and the cycle duration. Specifically, a log-normal distribution is setup by matching the inverse of cycle length 1/672 (1/h) and the absorption rate 2 (1/h) as the
0.025 and 0.975 quantiles, respectively. This gives a log-normal
distribution with the mean parameter -0.15. Hence, we assume $\text{log}(k_{\text{eff}})= -0.15$. The calculated $\text{E}(t|d,f)$ and $\text{AUC}_{E}(t|d,f)$ of the Vidaza trial for combinations of the 8 mg-Schedule D and the 24 mg-Schedule B 
are displayed in Figure 2. Notice that the $\text{AUC}_{E}(t|d,f)$ of the 24
mg-Schedule B at week 4 is 1, since this dose-schedule combination is taken as the
reference dose-schedule combination and the length of cycle 1 is 4 weeks. 

\begin{figure}
\centering
  \includegraphics[scale=0.15]{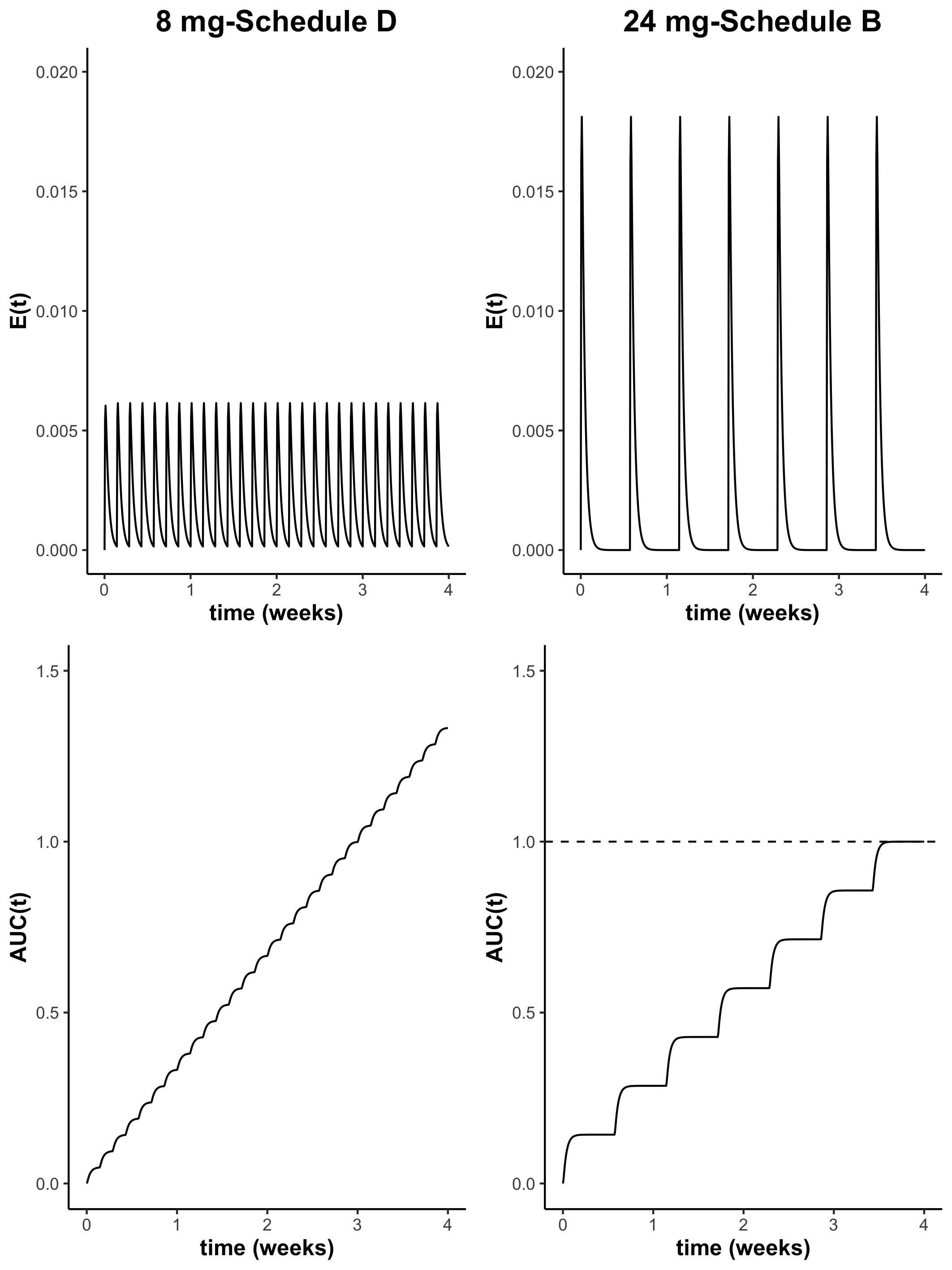}
  \caption{Illustration of the exposure measure of the drug ($\text{E}(t|d,f)$) and the AUC
    over exposure measure ($\text{AUC}_{E}(t|d,f)$) for the 8
    mg-Schedule D and the 24 mg-Schedule B combinations of the Vidaza trial,
    respectively. The reference dose-schedule combination is the 24 mg-Schedule B and the length of
    cycle 1 is 4 weeks.}
\label{fig:pseudoPK}
\end{figure}

For TITE-PK model, a normal weakly informative prior is chosen for log($\beta$) 
with a standard deviation of 1.75 and a mean which corresponds to $P(T \leq t^{*} | d^{*}, f^{*})$ of 0.3. We used true values of $k_{e}$ and $k_{\text{eff}}$ for the estimation. The comparison of the prior 
DLT probabilities used by TITE-PK and the prior skeletons 
used by POCRM are displayed in Figure A1 in Appendix A. 

We did not consider the method by Braun et al.\cite{Braun2007}
in the simulations, since this method mostly requires a 
mean of approximately 60 patients to be enrolled which is 
not practical for many phase I trials. POCRM assumes that DLT probabilities increase monotonically with dose within each schedule. For different schedules, it specifies multiple possible orderings of dose-schedule combinations and uses model selection techniques to select the most appropriate model.\cite{wages2017} POCRM with partially ordered schedules relaxes the assumption of completely ordered schedules. This is done by specifying appropriate possible orderings. In the simulations, two versions of the POCRM, one assuming complete order 
schedules and one with partial order schedules, are 
considered. Computations of the POCRM were carried out using the publicly available \texttt{R} code 
provided by Wages et al.\cite{Wages2014} (see 
\url{http://faculty.virginia.edu/model-based_dose-finding}). 
We refer to Wages et al.\cite{Wages2014} for more 
information about the POCRM.

In the simulations, data for 1,000 trials were generated per scenario. 
For all methods, patients were 
assigned one at a time until the trial was stopped, the MTC was identified or the maximum number of 
patients per trial of 60 were used. For the TITE-PK, if all dose-schedule combinations are in the overdosing interval based on the adapted EWOC criterion, the trial is stopped without selecting any combination as the MTC. Otherwise, the trial continues until the recommendation of the MTC. The recommended MTC must meet the following conditions:
\begin{enumerate}
\item[(i)] At least 9 patients have been treated at the MTC.
  \item[(ii)] The recommended MTC satisfies one of the following conditions:
  \begin{itemize}
  \item The probability of targeted toxicity at the MTC
    exceeds 50\%: $P(0.20 \leq P(T \leq t^{*}|d,f) \leq 0.40) \geq 0.50$.
    \item A minimum of 21 patients have already been treated in the
      trial.
  \end{itemize}
\end{enumerate}

We consider two values for the feasibility bound, 
namely $a=0.25$ and $a=0.50$. Increasing the 
feasibility bound results in the more aggressive dose-schedule 
decisions, that is increasing the percentage of trials 
with the MTC declared in the targeted toxicity interval, 
while increasing the percentage of trials with the MTC declared
in the overdosing interval.

\subsection{Results} \label{results}
\begin{table}
\centering
\caption{Simulation results for two applications of POCRM with complete and partial order schedules and two applications of the proposed method TITE-PK with a feasibility bound of $a=0.25$ and $a=0.50$.}
\label{t2:results}
\begin{tabular}{llllllll}
  \toprule
            &   \multicolumn{7}{c}{\textbf{Scenario}} \\ 
  \midrule
           & 1 & 2 & 3 & 4 & 5 & 6 & 7 \\
  \midrule  
            \multicolumn{8}{c}{Probability of selecting MTC in the targeted toxicity interval} \vspace{0.2cm}  \\
  POCRM (complete)   & 0.62 & n/a & 0.74  & 0.65 & 0.42  & 0.57 & 0.52         \\
  POCRM (partial)       & 0.65 & n/a & 0.74  & 0.63 & 0.44  & 0.66 & 0.57        \\
  TITE-PK (a = 0.25)              & 0.42 & n/a & 0.56  & 0.36 & 0.22  & 0.36 & 0.28           \\
  TITE-PK (a = 0.50)              & 0.72 & n/a & 0.79  & 0.55 & 0.42  & 0.74 & 0.57          \\
            \multicolumn{8}{c}{Probability of selecting MTC in the overdosing interval} \vspace{0.2cm}  \\
  POCRM (complete)   & 0.00 & 0.49  & 0.08  & 0.20 & 0.36 & 0.00 & 0.26        \\
  POCRM (partial)       & 0.00 & 0.50  & 0.10  & 0.21 & 0.36 & 0.00 & 0.26          \\
  TITE-PK (a = 0.25)              & 0.00 & 0.06  & 0.01  & 0.03 & 0.04 & 0.00 & 0.05           \\
  TITE-PK (a = 0.50)              & 0.00 & 0.28  & 0.08  & 0.16 & 0.16 & 0.00 & 0.19           \\
            \multicolumn{8}{c}{Probability of selecting no combination as MTC} \vspace{0.2cm}    \\
  POCRM (complete)   & 0.05 & 0.51& 0.03    & 0.03 & 0.01 & 0.06 & 0.09         \\
  POCRM (partial)       & 0.05 &  0.50&  0.03  & 0.03 & 0.01 & 0.02 & 0.10          \\
  TITE-PK (a = 0.25)              & 0.10 & 0.94  & 0.14  & 0.16 & 0.08 & 0.13 & 0.39           \\
  TITE-PK (a = 0.50)              & 0.03 & 0.72  & 0.02  & 0.03 & 0.02 & 0.04 & 0.10          \\
            \multicolumn{8}{c}{Mean number of patients enrolled in the overdosing interval} \vspace{0.2cm}    \\
  POCRM (complete)      & 0.0  & 17.0  & 2.5  & 6.1  & 9.3  & 0.0 & 8.1      \\
  POCRM (partial)          & 0.0  & 17.6  & 3.2  & 7.0  & 10.1 & 0.0 & 8.2        \\
  TITE-PK (a = 0.25)                & 0.0  & 1.0  & 2.7   & 3.6  & 4.8  & 0.0 & 4.0        \\
  TITE-PK (a = 0.50)                & 0.0  & 4.6  & 7.3   & 8.5  & 9.4  & 0.0 & 7.6          \\
  
            \multicolumn{8}{c}{Mean number of patients enrolled in total} \vspace{0.2cm}                   \\
  POCRM (complete)    & 25.6 & 17.0  & 24.5   & 24.2 & 24.6  & 25.2 & 22.2         \\
  POCRM (partial)        & 25.9 & 17.6  & 25.7   & 25.3 & 25.8  & 25.9 & 23.6         \\
  TITE-PK (a = 0.25)              & 21.4 & 3.6   &  18.7   & 18.2 & 18.8  & 20.9 & 13.7         \\
  TITE-PK (a = 0.50)              & 18.7 & 8.5   &  20.6  & 21.1 & 21.0  & 18.6 & 19.7         \\

            \multicolumn{8}{c}{Mean number of DLT observed} \vspace{0.2cm}                        \\
  POCRM (complete)   & 4.6 & 8.9  & 6.6  & 7.2 & 7.1  & 4.5 & 7.6                   \\
  POCRM (partial)       & 4.7 & 9.3  & 6.9  & 7.8 & 7.7  & 4.7 & 8.3                   \\
  TITE-PK (a = 0.25)              & 4.0 & 4.5  & 1.9  & 4.6 & 4.7  & 3.8 & 4.1                    \\
  TITE-PK (a = 0.50)              & 4.3 & 6.7  & 4.8  & 7.3 & 7.4  & 4.1 & 7.5                    \\
   \bottomrule
\end{tabular}
\end{table}

Table~\ref{t2:results} shows the summary statistics for the performance of the methods under the seven scenarios. In Scenario 1, TITE-PK with $a=0.50$ outperforms both POCRM methods in terms of recommending the MTC in the targeted toxicity interval. The corresponding percentages are 72\%, 62\% and 65\% for TITE-PK ($a=0.50$), POCRM (complete) and POCRM (partial), respectively. However, TITE-PK with $a=0.25$ yields the worst performance according to the same measure, the corresponding percentage is 42\%. Scenario 1 includes no dose-schedule combinations in the underdosing interval, hence probability of selecting no combination as the MTC is 0 for all methods. In Scenario 2, all combinations are in the overdosing interval. TITE-PK with $a=0.25$ and $a=0.50$ stop the trial without the MTC selection in 94\% and 72\% of the time, respectively. However, both POCRM methods stop the trial around 50\% of the time. Subsequently, the percentages of the MTC selection in the overdosing interval of POCRM methods are higher than both TITE-PK models. Moreover, TITE-PK with $a=0.50$ is superior to POCRM with partial ordering in terms of mean number of patients enrolled in the overdosing interval (8.5 vs 17.6) and mean number of DLT observed (4.8 vs 9.3). An advantage of TITE-PK is that, due to early stopping and a small mean number of patients, the design expose approximately half of the sample size to toxic combinations in comparison to PO-CRM with partial order (given above).

In Scenario 3, TITE-PK with $a=0.50$ gives higher percentage for recommending a combination in the targeted toxicity interval than both POCRM methods. The corresponding percentages are 79\% for TITE-PK ($a=0.50$), 74\% for POCRM (complete), and 74\% for POCRM (partial). TITE-PK with $a=0.25$ selects the MTC in the overdosing interval in about 1\% of the time for Scenario 3, while it yields the worst performance with 56\% in terms of recommending a combination in the targeted toxicity interval.  In Scenario 4, both POCRM methods yield higher percentages than TITE-PK with $a=0.50$ for the selection of the MTC in the targeted toxicity interval. The corresponding percentages are 55\%, 65\% and 63\% for TITE-PK ($a=0.50$), POCRM (complete) and POCRM (partial), respectively. However, both POCRM methods yield slightly higher percentages than TITE-PK with $a=0.50$ for the selection of the MTC in the overdosing interval. TITE-PK with $a=0.50$ recommends the MTC in the overdosing interval in 16\% of the trials, while POCRM with complete ordering and POCRM with partial ordering do this in 20\% and 21\% of the trials.

Scenario 5 needs special consideration, since all methods perform poorly in terms of the MTC selection in the targeted toxicity interval. Consistent with other scenarios, TITE-PK ($a=0.25$) results in the lowest probability for the MTC selection in the targeted toxicity interval. However, both TITE-PK methods display superior performance in terms of selecting MTC in the overdosing interval. The corresponding percentages are 16\% for TITE-PK ($a=0.50$) and 36\% for POCRM with partial ordering. Scenario 6 corresponds to Scenario 1 but with DLT probabilities for Schedules B and C switched. In Scenario 6, TITE-PK ($a=0.50$) displays better performance compared to POCRM methods as in Scenario 1. In Scenario 7, TITE-PK ($a=0.50$) yield lower percentages than POCRM methods in terms of selecting the MTC in the overdosing interval. Although Scenarios 6 and 7 assume partial orderings, our method performs relatively well showing robustness against the violation of the complete ordering assumption.

\begin{table}
\centering
\caption{Simulation results for schedules recommended by TITE-PK (a = 0.50) and POCRM (partial) 
as part of the MTC. Schedules A, B, C, and D have dosing frequency 
of 192, 96, 48, and 24 hours, respectively.}
\label{t3:schedule}

\begin{tabular}{llllllll}
  \toprule
            &   \multicolumn{7}{c}{\textbf{Scenario}} \\ 
  \midrule
             & {1}  & {2}   & {3}   & {4}  & {5}  & {6}  & {7} \\
  \midrule  
            \multicolumn{8}{c}{Probability of selecting Schedule A as part of MTC} \vspace{0.2cm}  \\
  TITE-PK     & 0.00 & 0.25  & 0.12  & 0.33 & 0.46 & 0.01 & 0.55   \\
  POCRM      & 0.02 & 0.45 & 0.17  & 0.28 & 0.16 & 0.02 & 0.36    \\
  
            \multicolumn{8}{c}{Probability of selecting Schedule B as part of MTC} \vspace{0.2cm}  \\
  TITE-PK    & 0.08 & 0.02  & 0.47  & 0.56 & 0.46 & 0.08 & 0.15   \\
  POCRM      & 0.19 & 0.04  & 0.30  & 0.34 & 0.45 & 0.27 & 0.20    \\
  
            \multicolumn{8}{c}{Probability of selecting Schedule C as part of MTC} \vspace{0.2cm}  \\
  TITE-PK      & 0.25 & 0.02  & 0.34  & 0.08 & 0.06 & 0.16 & 0.19   \\
  POCRM      & 0.23 & 0.01  & 0.29  & 0.29 & 0.27 & 0.15 & 0.33    \\

            \multicolumn{8}{c}{Probability of selecting Schedule D as part of MTC} \vspace{0.2cm}  \\
  TITE-PK      & 0.63 & 0.00  & 0.05  & 0.00 & 0.01 & 0.72 & 0.01   \\
  POCRM      & 0.51 & 0.00  & 0.20  & 0.06 & 0.10 & 0.50 & 0.02    \\

   \bottomrule
\end{tabular}

\end{table}

We also examined which schedules are recommended by TITE-PK ($a=0.50$) and POCRM with partial ordering as part of the MTC. These results are listed in Table~\ref{t3:schedule}. In many scenarios, the schedules recommended by POCRM are more spread across four schedules compared to the schedules recommended by TITE-PK. For example, in Scenario 5 the MTC selection in the targeted toxicity interval is almost same for two methods (the only combination in the targeted toxicity interval is in Schedule B). However, the selected schedules by two methods are quite different. This results in inferior performance of POCRM in terms of the MTC selection in the overdosing interval. This is because, Schedule A of Scenario 5 which was selected by TITE-PK in 46\% of the time has less toxic dose-schedule combinations than Schedules C and D. The fact that TITE-PK selects the schedules more precisely is reflected in its superior performance in terms of the MTC selection in the targeted toxicity and/or overdosing intervals. One possible reason is that the EWOC criterion used by TITE-PK does not allow to escalate to the schedules with higher toxicity in comparison to POCRM, hence improving the overall performance.

Overall, TITE-PK with $a=0.25$ yields more conservative behaviour in terms of the MTC selection in the targeted toxicity and the overdosing intervals in comparison to TITE-PK with $a=0.50$, as it is expected. In all scenarios, TITE-PK with $a=0.25$ does not select the MTC in the overdosing interval more than 11\% of the time. Furthermore, TITE-PK with $a=0.25$ induces the lowest number of DLT in all scenarios and enroll the lowest number of patients to the overdosing interval. However, this conservative behaviour consistently results in a weaker performance in terms of selecting the MTC in the targeted toxicity interval. The main reason of this poor behaviour is related to the EWOC criterion and the choice of $a=0.25$. TITE-PK with $a=0.50$ yields superior or similar performance compared to the POCRM methods in terms of selecting the MTC in the targeted toxicity interval with the exception of Scenario 4. In terms of the MTC selection in the overdosing interval, TITE-PK with $a=0.50$ performs consistently better than POCRM methods. Furthermore, POCRM (partial) does not display clear benefit over TITE-PK with $a=0.50$ for Scenario 6 and 7 in which the assumption of complete ordering is relaxed. Finally, both TITE-PK models enroll lower number of patients compared to both POCRM methods in all scenarios. One reason of the desirable performance of TITE-PK may stem from the use of EWOC criterion, which reduces the risk of recommending toxic dose-schedule combinations as the MTC.

With the proposed method TITE-PK, time-to-DLT is modeled using a non-homogeneous Poisson distribution which has been used to simulate the timing of the events in the previously discussed scenarios. To examine the robustness of the Poisson process assumption following the exposure metric in TITE-PK, we generated datasets from different time-to-DLT models, namely uniform and exponential distributions, under each scenario. Furthermore, we considered a third data-generating process, that is assuming time-to-DLT occurring with higher probability at the early (between 0 and $\frac{t^{*}}{5}$) or late (between time $\frac{4\,t^{*}}{5}$ and $t^{*}$) stages within the first cycle.
For the uniform distribution, the occurrences of DLT are determined using the true DLT probabilities. The timing of DLT is sampled uniformly within the first cycle. For the exponential distribution, the rate parameter of the exponential distribution is calculated by $\lambda = -\text{log}(1-P(T \leq t^{*} | d, f))/t^{*}$ where $P(T \leq t^{*} | d, f)$ corresponds to a true DLT probability.\cite{MGB} Then, the timing of DLT is sampled using the exponential distribution with the specified rate parameter within the first cycle. For the third data-generating process, the occurrences of DLT are determined using the true DLT probabilities as in the uniform distribution. The timing of DLT is sampled assuming with the probability of 0.4 for the interval 0 and $\frac{t^{*}}{5}$ (early) and the probability of 0.4 for the interval $\frac{4\,t^{*}}{5}$ and $t^{*}$ (late). These give that within the interval $\frac{t^{*}}{5}$ and $\frac{4\,t^{*}}{5}$, the corresponding probability is 0.2. We only considered the feasibility bound of $a=0.50$. Table~\ref{t4:robust} gives the results of three performance measures. The first rows of each performance measure replicate the values displayed in Table~\ref{t2:results} of TITE-PK ($a=0.50$) values. Table~\ref{t4:robust} indicates that the performance of TITE-PK varies little with the time-to-DLT distribution in terms of investigated measures.

\begin{table}
\centering
\caption{Simulation results under different time-to-DLT distributions: TITE-PK, uniform and exponential distributions, and time-to-DLT occurring with higher probability at the early (between time 0 and $\frac{t^{*}}{5}$) or late (between time $\frac{4\,t^{*}}{5}$ and $t^{*}$) stage within the first cycle.}
\label{t4:robust}
\begin{tabular}{llllllll}
  \toprule
            &   \multicolumn{7}{c}{\textbf{Scenario}} \\ 
  \midrule
           & 1 & 2 & 3 & 4 & 5 & 6 & 7 \\
  \midrule  
            \multicolumn{8}{c}{Probability of selecting MTC in the targeted toxicity interval} \vspace{0.2cm}  \\
  TITE-PK      & 0.72 & 0.00  & 0.79  & 0.55 & 0.42 & 0.74 & 0.57   \\
  Uniform      & 0.76 & 0.00  & 0.79  & 0.58 & 0.42 & 0.76 & 0.54    \\
  Exponential  & 0.75 & 0.00  & 0.80  & 0.56 & 0.46 & 0.74 & 0.54    \\
  Early/late & 0.76 & 0.00 & 0.80 & 0.57 & 0.42 & 0.76 & 0.54      \\
  
            \multicolumn{8}{c}{Mean number of patients enrolled in the overdosing interval} \vspace{0.2cm}    \\
  TITE-PK      & 0.0 & 8.5  & 4.6  & 7.3 & 9.4  & 0.0 & 7.6           \\
  Uniform      & 0.0 & 9.9 & 5.1  & 8.6 & 10.4 & 0.0 & 8.7  \\
  Exponential  & 0.0 & 9.6  & 4.7  & 8.0 & 10.1 & 0.0 & 8.4   \\
  Early/late   & 0.0 & 9.6 & 4.8 & 8.2 & 10.0 & 0.0 & 8.9    \\
  
            \multicolumn{8}{c}{Mean number of patients enrolled in total} \vspace{0.2cm}        \\
  TITE-PK      & 18.7 & 8.5   & 20.6  & 21.1 & 21.0 & 18.6 & 19.7    \\
  Uniform      & 18.4 & 9.9   & 20.4  & 21.0 & 21.1 & 18.2 & 19.8    \\
  Exponential  & 18.9 & 9.6   & 20.4  & 21.4 & 21.5 & 18.6 & 19.5   \\
  Early/late    & 19.0 & 9.6  & 20.0 & 20.8 & 21.0 & 18.7 & 19.8   \\

   \bottomrule
\end{tabular}
\end{table}

\section{Discussion and Conclusions} \label{sec:discuss}
We propose a Bayesian adaptive model, a time-to-event pharmacokinetic (TITE-PK) model, to 
support design and analysis of phase I dose-escalation trials with 
multiple schedules to provide guidance for the dose-schedule decisions. TITE-PK has an
interpretable parameter which facilitates the prior specification. It uses pharmacokinetic (PK) principles
to combine different treatment schedules in a
model-based approach. An adapted escalation with overdose
control (EWOC) criterion can be used with TITE-PK.
In the simulations, for six of seven scenarios considered, 
TITE-PK with a feasibility bound of 0.50 shows superior or similar 
performance in terms of identifying the maximum tolerated dose-schedule combination (MTC) in the targeted toxicity interval 
compared to the partial order continual reassessment method (POCRM). In terms of the recommendation of the MTC in
the overdosing interval, TITE-PK yields lower percentages in
all seven scenarios considered. For all scenarios, the TITE-PK model 
required lower numbers of patients enrolled compared to POCRM.

Here, we considered simultaneously finding a suitable dose-schedule
combination within a phase I trial as in Wages et al.\cite{Wages2014} 
Another useful design would investigate multiple schedules, say Schedules 1 and 2, sequentially. That is, 
enrolling dose cohorts of patients with Schedule 1 and
estimating the maximum tolerated dose (MTD) for the Schedule 1. Then, patients are 
enrolled into dose cohorts using Schedule 2 and the MTD is estimated
for Schedule 2 by utilizing data coming from both 
schedules. TITE-PK can be used to design
such sequential phase I trials or a phase I trial involving only one schedule.
As such designs are beyond the scope of this paper, they are not investigated here.

Here, we considered the 
cohort size of 1. However, there is no restriction 
in TITE-PK regarding the cohort size. We defined the schedule as the frequency of administration within a cycle. Nevertheless, TITE-PK can be used to design trials with the other definition of the schedule, that is the number of cycles to treat patients. This can be achieved by assigning different reference time point $t^{*}$ for different dose-schedule combinations based on the number of treatment cycles. Frequency of administrations $f$ for different dose-schedule combinations will be assumed to be the same.

One limitation of the proposed method is the assumption of monotonicity of the
exposure-DLT probability relationship. Moreover, this 
monotonicity assumption implies a monotonic dose-DLT curve, since a linear PK is 
used in the pseudo-PK model. Violation of the linear PK assumption can be informed
using the external PK data from the ongoing trial.
To relax the assumption of linear PK, one can consider more
complicated PK models including a non-linear PK model which may not
have an analytical solution. Such extensions may be 
implemented in \texttt{Stan} which 
has a built-in differential equation solver. However, 
more complicated modelling approaches always need
to be calibrated well given the sparseness of the phase I
dose-escalation data sets. Alternatively, one can consider an ad-hoc
extension of the TITE-PK model. For instance, by introducing a non-linearity factor $\gamma$,\cite{doi:10.1002/sim.4780101104} a \emph{pseudo}-dose as $(\frac{d}{d^{*}})^{\gamma}$ can be used
instead of dose $d$ in the model which may be helpful to relax the linear PK assumption. 

When relevant historical information or data from a different study
population exists, it is desirable to include such information in the
analysis of the phase I trial, for example using a meta-analytic-predictive (MAP) prior.\cite{BIOM:BIOM12242} TITE-PK can be extended to use a MAP approach. 
Another crucial aspect of the methods for phase I trials is the
ability to analyse the combination of drugs. Although we only
consider the single agent case here, it is possible to extend TITE-PK
to analyse drug combinations which is complicated by the need to model
possible drug interactions. Another 
extension of TITE-PK is considering a two-parameter version in 
which one of the PK parameters $k_{\text{eff}}$ 
is also estimated in the model jointly with the regression coefficient 
$\beta$.

\section*{Acknowledgements}
We thank Heinz Schmidli who pointed
us to several important references, Michael Looby for 
proofreading an earlier version of the manuscript, Abdelkader Seroutou and Christian R\"over for contributing valuable comments. We thank the Associate Editor and two 
anonymous reviewers whose comments and suggestions helped to improve this paper.

{
\appendix
\section{Prior probabilities for TITE-PK and prior skeletons for POCRM}\label{app1}

Figure A1 shows the comparison of the prior 
DLT probabilities used by TITE-PK and the prior skeletons 
used by POCRM. 

\begin{figure}
  \centering
  \includegraphics[scale=0.2]{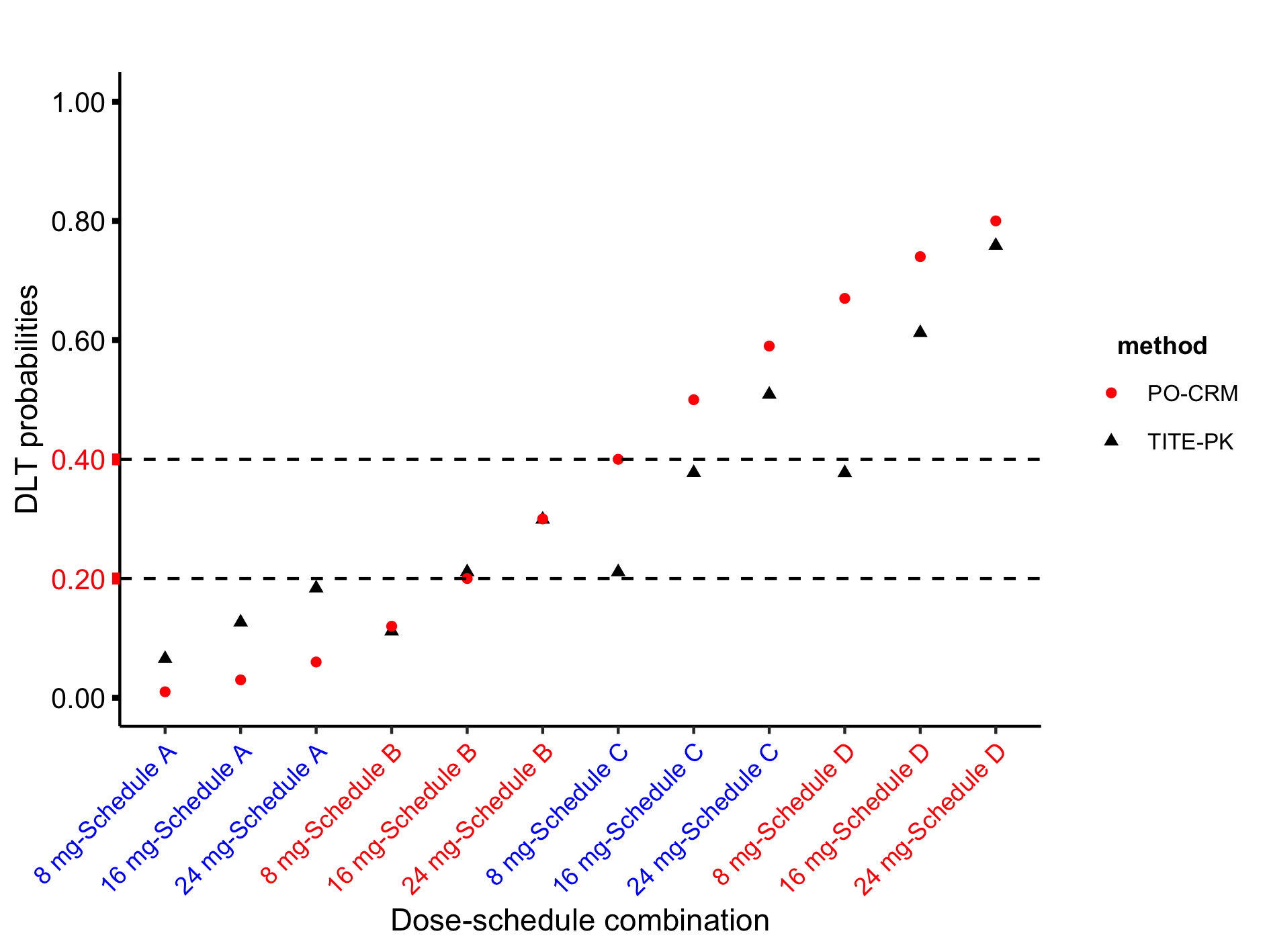}
  \caption{Prior DLT probabilities obtained by TITE-PK (as triangles) and the prior skeletons used by POCRM (as circles). The horizontal dashed lines represents the boundaries of the targeted toxicity interval. Schedules A, B, C, and D have dosing frequency 
of 192, 96, 48, and 24 hours, respectively.}
  \label{fig:Priors}
\end{figure}

\section{Additional simulation results}\label{app2}
We also conducted simulations to investigate further toxicity scenarios. The scenarios and the results are illustrated in Table~\ref{tab:app1} and Table~\ref{tab:app2}, respectively.

\begin{table}
\centering

\caption{Additional simulation scenarios: Toxicity scenarios for the dose-schedule combination in the simulation study. Combinations in the targeted toxicity interval (0.20 - 0.40) are in boldface. Schedules A, B, C, and D have dosing frequency 
of 192, 96, 48, and 24 hours, respectively.}
\label{tab:app1}

\begin{tabular}{lllllll}
  \toprule
             & \multicolumn{6}{c}{\textbf{Doses in mg/$\text{m}^2$}} \\
             \cmidrule{2-7}
   Schedule  & 8 & 16 & 24 & 8 & 16 & 24 \\ 
  \midrule
             & \multicolumn{3}{c}{Scenario 8} & \multicolumn{3}{c}{Scenario 9} \\
   A & 0.10 & 0.26 & 0.35 & 0.10 & 0.28 & 0.45  \\
   B & 0.30 & 0.32 & 0.50 & 0.12 & \textbf{0.30} & 0.48 \\
   C & 0.45 & 0.50 & 0.62  & 0.14 & \textbf{0.32} & 0.55 \\
   D & 0.55 & 0.62 & 0.72 & \textbf{0.30} & 0.48 & 0.70 \\ 
   
                &  \multicolumn{3}{c}{Scenario 10} \\
   A  & 0.01 & 0.10 & 0.50\\
   B & 0.05 & 0.50 & 0.60 \\
   C & 0.03 & \textbf{0.30} & 0.55 \\
   D & 0.10 & 0.60 & 0.70 \\

   \bottomrule

\end{tabular}

\end{table}

\clearpage

\begin{table}
\centering

\caption{Additional simulation results of POCRM with partial order schedules and the proposed method TITE-PK with a feasibility bound of $a=0.50$.}
\label{tab:app2}

\begin{tabular}{llll}

  \toprule
            &   \multicolumn{3}{c}{\textbf{Scenario}} \\ 
  \midrule
           & 8 & 9 & 10 \\
  \midrule  
            \multicolumn{4}{c}{Probability of selecting MTC in the targeted toxicity interval} \vspace{0.2cm}  \\
  POCRM (partial)       & 0.62 &  0.63 &   0.44        \\
  TITE-PK (a = 0.50)    & 0.68 &  0.75 &   0.21           \\
            \multicolumn{4}{c}{Probability of selecting MTC in the overdosing interval} \vspace{0.2cm}  \\
  POCRM (partial)       & 0.22 & 0.17  & 0.33          \\
  TITE-PK (a = 0.50)   & 0.11 & 0.12 & 0.30           \\
            \multicolumn{4}{c}{Probability of selecting no combination as MTC} \vspace{0.2cm}    \\
  POCRM (partial)       & 0.10 & 0.10  & 0.01            \\
  TITE-PK (a = 0.25)   & 0.11 & 0.08 & 0.01             \\
            \multicolumn{4}{c}{Mean number of patients enrolled in the overdosing interval} \vspace{0.2cm}    \\
  POCRM (partial)        & 7.9  & 5.7  & 10.0         \\
  TITE-PK (a = 0.50)    & 6.1  & 6.8  & 11.3            \\
  
            \multicolumn{4}{c}{Mean number of patients enrolled in total} \vspace{0.2cm}                   \\
  POCRM (partial)       & 24.0 & 24.8  & 25.7           \\
  TITE-PK (a = 0.50)  & 19.1 & 19.3  &  21.6           \\

            \multicolumn{4}{c}{Mean number of DLT observed} \vspace{0.2cm}                        \\
  POCRM (partial)       & 8.4 & 7.4 & 7.7                     \\
  TITE-PK (a = 0.50)  & 7.0 & 7.0  & 7.9                    \\
   \bottomrule
   
\end{tabular}

\end{table}

}
\vfill

\clearpage

\bibliography{bibliography}

\end{document}